\documentclass[10pt,prl,aps,twocolumn,showpacs,floatfix]{revtex4}
\usepackage{graphicx}
\usepackage{amssymb}
\usepackage{amsmath}

\begin{document}

\title{Quantum versus classical annealing---insights from scaling theory \\
and results for spin glasses  on 3-regular graphs}

\author{Cheng-Wei Liu, Anatoli Polkovnikov, and Anders W. Sandvik}
\affiliation{Department of Physics, Boston University, 590 Commonwealth Avenue, Boston, Massachusetts 02215, USA}

\begin{abstract}
We discuss an Ising spin glass where each $S=1/2$ spin is coupled antiferromagnetically to three other spins (3-regular graphs). Inducing quantum 
fluctuations by a time-dependent transverse field, we use out-of-equilibrium quantum Monte Carlo simulations to study dynamic scaling at the quantum 
glass transition. Comparing the dynamic exponent and other critical exponents with those of the classical (temperature-driven) transition, we conclude 
that quantum annealing is less efficient than classical simulated annealing in bringing the system into the glass phase. Quantum computing based on 
the quantum annealing paradigm is therefore inferior to classical simulated annealing for this class of problems. We also comment on previous simulations 
where a parameter is changed with the simulation time, which is very different from the true Hamiltonian dynamics simulated here.
\end{abstract}

\date{\today}

\pacs{03.67.Ac, 05.30.Rt, 75.10.Jm, 75.50.Lk}

\maketitle

Simulated annealing (SA), which was first proposed in the context of spin glasses \cite{kirkpatrick83}, is one of the most versatile 
optimization methods \cite{cerny85,granville94}. The basic idea of SA is that a Monte Carlo (MC) simulation with slowly decreasing temperature 
can explore the energy (cost-function) landscape of a complex system without getting trapped in local minimums if the process is sufficiently 
slow (in analogy with removal of crystal defects by heating and annealing). It is natural to ponder the feasibility of similar schemes based on 
slow reduction of quantum fluctuations in {\it quantum annealing} (QA) processes. Such schemes have been explored for some time, theoretically
 \cite{finnila94,kadowaki98,santoro02} as well as in experiments on frustrated Ising systems such as LiHo$_x$Y$_{1-x}$F$_4$ 
\cite{bitko96,brooke99}. The QA ideas have risen to particular prominence in the context of quantum computation \cite{farhi01,hogg03,das08}, 
where there are now serious efforts to implement QA (also called the quantum-adiabatic algorithm) in actual devices \cite{dwave11}, 
currently with  $\approx 500$ q-bits in the D-Wave device \cite{dwave13}. It is not yet clear whether true QA has been realized in these 
systems, however \cite{boixo13,smolin14}. Beyond this practical issue, a fundamental question is whether QA really is more efficient 
than SA for important optimization problems. This question has been addressed \cite{altshuler09,bapst13} but so far there are 
few solid conclusions. 

We here present a generic way to compare SA and QA, using scaling theory in combination with a quantum MC (QMC) algorithm to simulate 
systems out of equilibrium with Hamiltonian dynamics in imaginary time \cite{degrandi11,liu13}.
We also show that this is very different from the dynamics arising when a parameter is changed versus QMC simulation 
time, as done in recent attempts to model a QA device \cite{boixo13,smolin14}.
We present results for an essential model studied in the context of QA; a quantum 
$S=1/2$ spin model on random 3-regular graphs in which all spins interact antiferromagnetically with three other spins. 
The corresponding classical Ising glass has an exactly known transition temperature and critical exponents \cite{zdeborov08,krazakala08}. 
The quantum model includes a transverse field and has a ground-state glass transition. Recent work has shown 
evidence for a continuous transition but the results were not completely conclusive \cite{farhi12}. Here we demonstrate a
continuous transition by scaling QMC data as a function of the velocity in the imaginary-time QA scheme. 
The exponents governing the critical growth of glass domains show that the QA is less efficient than the 
corresponding SA protocol. Thus, for a large system, a quantum computer based on the QA would not pass through 
the glass transition faster than a classical SA process.

{\it Quantum annealing.}---Many optimization problems can be cast in the form of energy minimization of a classical 
Ising spin system described by the Hamiltonian
\begin{equation}
H_0 = \sum_{i=1}^N\sum_{j=1}^N J_{ij} \sigma^z_i\sigma^z_j,~~~~~(\sigma_i^z = \pm 1).
\label{h0}
\end{equation}
Challenging problems correspond to disordered frustrated interactions $J_{ij}$.
In QA, quantum fluctuations of some form are added, e.g., a uniform transverse field
\begin{equation}
H_1 = h\sum_{i=1}^N \sigma_i^x = h\sum_{i=1}^N (\sigma_i^+ + \sigma_i^-).
\label{h1}
\end{equation}
The total Hamiltonian is expressed as
\begin{equation}
H=sH_0+(1-s)H_1,
\label{hs}
\end{equation}
where $s \in [0,1]$ regulates the quantum fluctuations. The ``driver'' $H_1$ can be chosen such that its ground state 
is trivial; with Eq.~(\ref{h1}) it is the product state $|\Psi_0(0)\rangle = \prod_i |\uparrow_i + \downarrow_i\rangle$. 
By the adiabatic theorem \cite{kato51,messiah62}, if the change of $s$ from $0$ to $1$ is sufficiently slow, the system will stay 
in the ground state $|\Psi_0(s)\rangle$ and in the limit $s\to 1$ one obtains an optimal 
solution (out of typically a large number of degenerate ones) of the classical problem.

The critical issue is how slowly $s$ must change for the solution not to be ruined by excitations. In the 2-level Landau-Zener problem 
the time is $\propto \Delta^{-2}$, where $\Delta$ is the minimum gap between the two states. Generalizing this to a many-body system with $N$ 
degrees of freedom, such as Eq.~(\ref{hs}), for large $N$ a quantum phase transition is expected at some point $s_c \in [0,1]$ where the ground 
state changes from trivial, in some sense, to complex. If at $s_c$ the gap is $\Delta_N$ and if $s$ is changed linearly, the required annealing 
time  grows with $N$ as $\Delta^{-p}_N$ \cite{farhi02} (though the claim $p \ge 2$ is inaccurate, as we will discuss below).
Then, if $\Delta_N \to 0$ as a power of $1/N$ (in a continuous quantum phase transition) one can solve the problem using QA in polynomial time 
in $N$. For an exponentially vanishing gap (first-order transition) the time grows exponentially. 

Arguments such as these have stimulated interest in numerically investigating quantum phase transitions in 
important quantum information problems. Initial results for one class of problems indicated a continuous transition \cite{farhi01,hogg03}, but 
once results for larger systems became available a first-order transition seemed more likely \cite{young08,young10}. Other problems have been investigated 
recently \cite{farhi12} and some of them likely have continuous transitions. 

An important issue was neglected above:
The nature of the quantum state and excitations once the critical point has been passed. While in models based on Eqs.~(\ref{h0}) and (\ref{h1})
the lowest excitations are gapped for $s < s_c$, the glassy state for $s > s_c$ should in general have dense gapless excitations. 
Therefore, going through the critical point is only the first stage of difficulties, and advancing further on the way to $s=1$ may be exponentially 
hard even for a power-law closing of the gap at $s_c$ \cite{santoro02,farhi12}. 
Nevertheless, the initial passage through the transition is clearly an important 
step to understand and quantify. Here we obtain insights and quantitative results based on scaling 
properties of the quantum and classical glass transitions in antiferromagnets on 3-regular graphs.

{\it Non-equilibrium QMC.}---One reason for the currently rather poor general understanding of the efficiency of QA schemes is the difficulties 
of studying dynamics of large quantum many-body systems on classical computers. Recently QMC simulations realizing Schr\"odinger evolution 
in imaginary time were proposed as a way to obtain limited but valuable information \cite{degrandi11,degrandi13,liu13}.
Here we use the {\it quasi-adiabatic QMC} (QAQMC) method \cite{liu13}, where $|\Psi_0(s=0)\rangle$ is acted upon by a product of 
$m$ evolving Hamiltonians $P_{m,1}=H(s_m)\cdots H(s_2)H(s_1)$, where in the simplest (linear quench) case $s_j = j\delta_s$ with $\delta_s=s_m/m$. 
The normalization $\langle \Psi_0(0)|P_{1,m}P_{m,1}|\Psi_0(0)\rangle$ is written as a sum over all possible strings of the operators 
in the terms (\ref{h0}) and (\ref{h1}), and ``asymmetric expectation values'' of the form
\begin{equation}
\langle A\rangle_\tau = \frac{\langle \Psi_0(0)|P_{1,m}P_{m,\tau+1}AP_{\tau,1}|\Psi_0(0)}{\langle \Psi_0(0)|P_{1,m}P_{m,1}|\Psi_0(0)},
\label{ataudef}
\end{equation}
are MC evaluated. The quantity $\langle A\rangle_\tau$ approaches the ground state expectation value $\langle A(s_\tau)\rangle$ when $m \to \infty$ and for 
finite $m$ it contains the same leading finite-velocity correction as in imaginary-time Schr\"odinger dynamics with an evolving Hamiltonian $H[s(t)]$, with 
$s= s_c - v(t_{\rm f}-t)$, $t \in [t_i,t_f]$, and the velocity $v = ds(t)/dt = aN\delta_s$. The factor $a$ is known \cite{liu13} but is irrelevant 
for scaling, and we here use $a=1$. Since imaginary- and real-time quenches to critical points share the same dynamic exponent 
$z$ \cite{degrandi11}, real-time critical scaling behavior can be extracted using QAQMC. We can also continue past $s_c$ into the glass 
phase but here our main aim is to study the dynamic criticality upon approaching $s_c$.

The implementation of the QAQMC method for the 3-regular graphs is a straight-forward generalization of the method developed for the standard 
transverse-field Ising model (TFIM) in Ref.~\onlinecite{liu13}. The classical part of the Hamiltonian is Eq.~(\ref{h0}), with any given spin $i$ 
coupled to exactly three other spins $j$, and for these pairs $J_{ij}=1$ (antiferromagnetic). The random graphs were generated 
using the Steger-Wormald algorithm \cite{steger99}. 

The physical quantity of main interest is the Edwards-Anderson spin-glass order parameter $q$, 
which is defined using two replicas (independent simulations), $1$ and $2$, of a given disorder realization of the random couplings;
\begin{equation}
q = \frac{1}{N}\sum_{i=1}^N \sigma^z_i(1)\sigma^z_i(2).
\label{qdef}
\end{equation}
We will analyze $\langle q^2\rangle$ averaged over
thousands of quenches of systems with different random couplings. As an illustration of dynamic scaling and different types of dynamics 
we will also study a ferromagnet, $J_{ij}=-1$ for all nearest-neighbor pairs ($j=i+1$) on a periodic chain. In this case we calculate the 
standard magnetization $m_z=(1/N)\sum_i \sigma^z_i$ and analyze $\langle m_z^2\rangle$. 

{\it Dynamic scaling.}---We will analyze data from QAQMC simulations within the framework of the Kibble-Zurek (KZ) scaling ansatz 
\cite{kibble76,zurek85} and its later 
generalizations \cite{polokovnikov05,zurek05,dziarmaga05,degrandi10,dziarmaga10,polkovnikov11,liu14}. 
The key point here is that there is a velocity $v_{\rm KZ}$ separating adiabatic and non-adiabatic evolution, and for a
system of length $L$ this is given by
\begin{equation}
v_{\rm KZ} \propto L^{-(zr+1/\nu)} \propto N^{-(z'r+1/\nu')},
\label{vc}
\end{equation}
where $\nu$ is the equilibrium exponent governing the divergence of the correlation length, $z$ is the dynamic exponent, and we 
have also introduced exponents normalized by the dimensionality $d$; $N=L^d$, $\nu'= \nu d$ and $z' = z/d$. The 3-regular graphs have $d=\infty$ 
and we will use $N$ for the size. To convert to unprimed exponents the upper critical dimension should then be used; $d=d_u$. 

The existence of a characteristic velocity suggests a generalized finite-size scaling form for singular quantities at
the critical point. For quantities calculated at the final time $t_f$ when $s=s_c$, and when $v \propto v_{\rm KZ}$ or lower, the
order parameter takes the form
\begin{equation}
\langle q^2\rangle \sim N^{-2\beta/\nu'}f(vN^{z'r+1/\nu'}),
\label{q2scalingform}
\end{equation}
and we can extract the important exponent combinations $\beta/\nu'$ and $z'+1/\nu'$ using a data-collapse technique with 
results for different $N$ and $v$ \cite{liu14}. In all cases discussed below, the resulting exponents are stable and  
insensitive to details of the fitting  procedures. We also note that the scaling form should work only at a continuous transition and 
its applicability, thus, supports such a transition.

{\it Hamiltonian versus simulation dynamics.}---Before presenting QAQMC results for the 3-regular graphs, let us comment on the method 
of changing $H$ as a function of the simulation time (instead of the imaginary-time evolution that we advocate). This approach 
is normally considered with thermal QMC simulations \cite{santoro02,santoro06,bapst13} but can also be implemented for QAQMC. 
To illustrate this we use the ferromagnetic $d=1$ TFIM. We use a relatively large number of operators in the operator sequence 
in (\ref{ataudef}), $m=4N^2$ (sufficient for ground-state convergence at all $s$ in equilibrium), and keep $s$ the same for all 
operators. The simulation starts at $s=0$ and $s$ is changed linearly at velocity $v$ until $s_c=1/2$ is reached. At this stage the magnetization 
is calculated. The procedure is repeated many times to obtain $\langle m_z^2\rangle$. The velocity 
is defined using a time unit of a sweep of either local updates (a Metropolis procedure where small segments of spins are flipped) or 
cluster updates (a generalization of the Swendsen-Wang, SW, cluster updates \cite{rieger94,sandvik03}) throughout the system. 

Applying the scaling ansatz in Eq.~(\ref{q2scalingform}) to $\langle m_z^2\rangle$, we extract the dynamic exponent characterizing 
the approach to the critical point when using local or cluster updates. We compare with the exponent obtained with the QAQMC simulations, 
where $s$ evolves within the operator string in Eq.~(\ref{ataudef}). In the latter case there is no dependence on the type of MC updates 
(but cluster updates give results with smaller statistical errors for a given simulation time) and we should detect Hamiltonian dynamics with $z=1$.

\begin{figure}
\centerline{\includegraphics[width=7.5cm, clip]{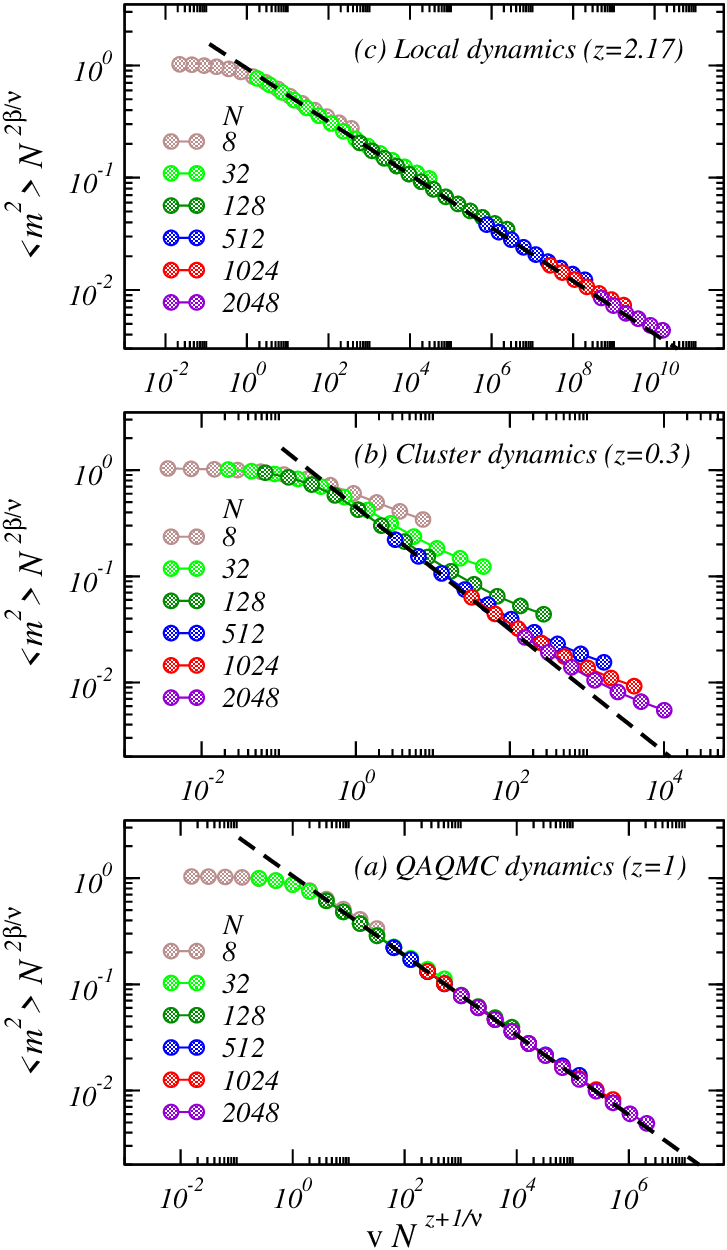}}
\vskip-2mm
\caption{(Color online) Velocity scaling for linear quenches of the TFIM: (a) Quantum quench with QAQMC Hamiltonian dynamics 
in imaginary time, (b) simulation-time quenches with Metropolis dynamics, and (c) SW cluster dynamics. The observed deviations from
the common scaling functions are expected at an $N$-dependent high velocity \cite{liu14}.}
\label{fig1}
\vskip-3mm
\end{figure}

The scaling analysis for all the cases is presented in Fig.~\ref{fig1}. The static exponents are those of the $d=2$ classical Ising model, 
$\beta=1/8$ and $\nu=1$, and we use these to produce scaling plots according to the form (\ref{q2scalingform}). We suspect that the simulation-time 
dynamics should be the same as in the classical $d=2$ Ising model with local and SW updates, and therefore test scaling with $z= 2.17$ and 
$z= 0.30$, respectively (as recently computed using KZ scaling in Ref.~\onlinecite{liu14}). The data collapse is very good in all cases for 
sufficiently large systems and low velocities. The lines in the log-log plots have slopes given by 
\begin{equation}
x=\frac{d-2\beta/\nu}{zr+1/\nu} =\frac{1-2\beta/\nu'}{z'r+1/\nu'},
\label{xdef}
\end{equation}
for $v_{KZ} \alt v \ll 1$ \cite{liu14}. For $v \approx v_{\rm KZ}$ there is a cross-over to equilibrium finite-size 
scaling, where $\langle m_z^2\rangle \propto N^{-2\beta/\nu}$. For $v$ of order $1$ there is high-velocity cross-over (not clearly 
seen in Fig.~\ref{fig1}) into a size-independent $\langle m_z^2\rangle$, governed by another scaling form \cite{liu14}. 

The above 
results for the dynamic exponents obtained under different evolution schemes confirm that evolving a model in simulation time does not access 
Hamiltonian dynamics and has little relevance for studying QA. While we have here explicitly demonstrated this in the case of dynamic
critical scaling, there is also no reason to expect the stochastic simulation-time dynamics to be relevant to quantum evolution in 
the glass phase. Hence the conclusions on the quantum mechanical nature of the dynamics of the D-Wave device drawn in 
Ref.~\onlinecite{boixo13} on the basis of such calculations are questionable (see also Ref.~\cite{smolin14}).

\begin{figure}
\centerline{\includegraphics[width=8.4cm, clip]{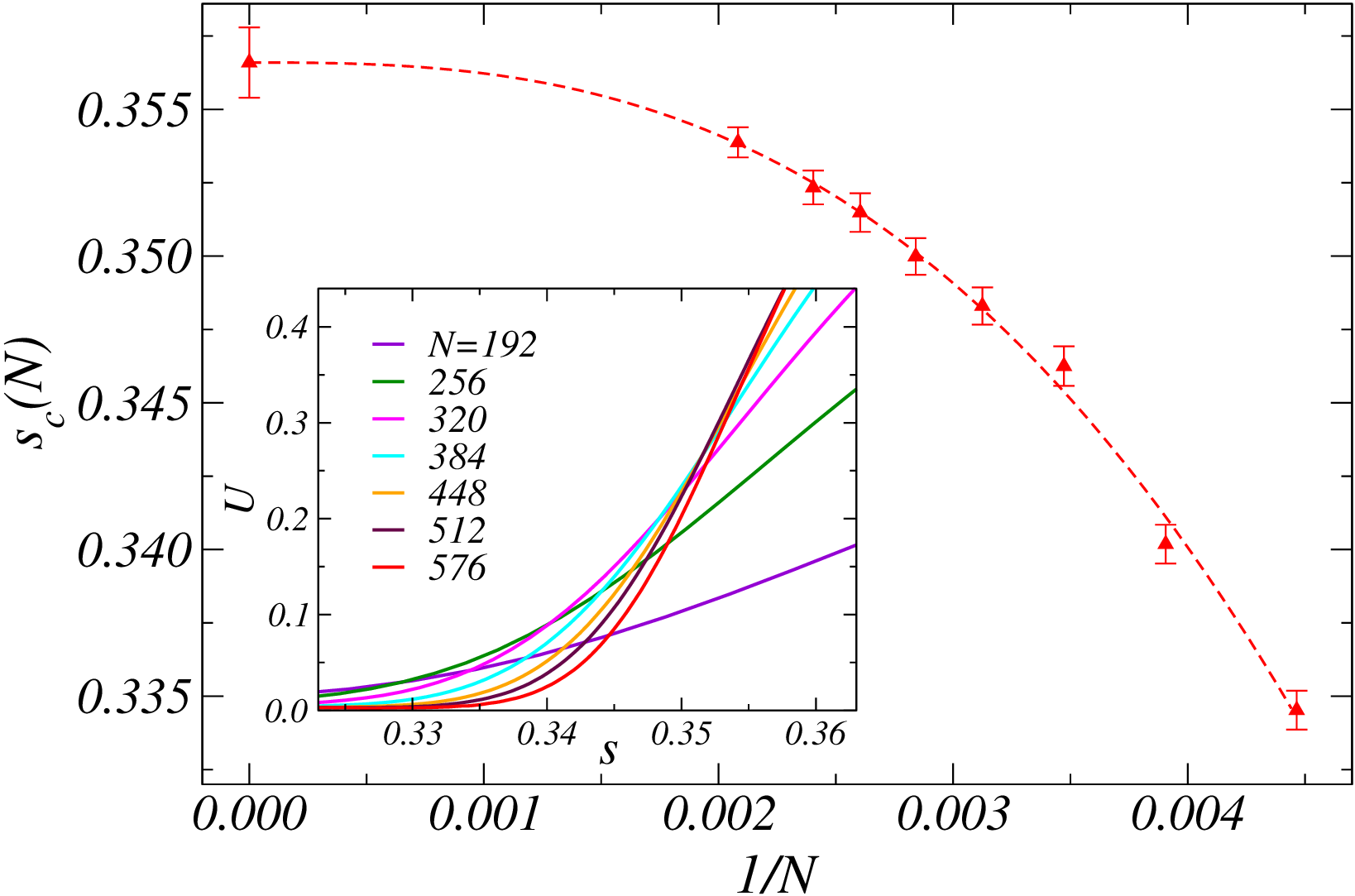}}
\vskip-2mm
\caption{(Color online) 
Crossing points between Binder cumulants for 3-regular graphs with $N$ and $N+64$ spins, extracted using the curves shown in the inset. 
The results were obtained in quenches with $v \sim N^{-\alpha}$ for $\alpha=17/12$. The curve in the main panel is a power-law fit 
for extrapolating $s_c$.}
\label{fig2}
\vskip-3mm
\end{figure}

{\it QA on 3-regular graphs.}---For the classical antiferromagnetic 3-regular graphs $T_c=-2\ln^{-1}[1-2/(1+\sqrt{2})]$ and the exponents, 
including $z$ for SA with local updates, are also known; $\beta=1$, $\nu'=3$, $z'=2/3$ ($d_u=6$) \cite{zdeborov08,krazakala08}. 
We have tested the scaling approach on this system and reproduced 
$T_c$ and the exponents to within a few percent \cite{longpaper}. Adding the transverse field (\ref{h1}), based on the quantum cavity 
method a value $s_c \approx 0.37$ was found in Ref.~\onlinecite{farhi12}, and QMC calculations of excitation gaps were in good agreement 
with this estimate. The expected errors in these calculations are of order several percent. 

We have located $s_c$ using QAQMC with $v \propto N^{-\alpha}$, where $\alpha$ exceeds the KZ exponent 
$z'+1/\nu'$ (which is unknown but later computable for {\it a posteriori} verification). Then $\langle q^2\rangle \sim N^{-2\beta/\nu'}$ at $s_c$
because $f(x)$ in Eq.~(\ref{q2scalingform}) approaches a constant when $x \to 0$. As illustrated in Fig.~\ref{fig2}, quenching past the estimated $s_c$,
we use a curve crossing analysis of the Binder cumulant, $U=(3-\langle q^4\rangle/\langle q^2\rangle^2)/2$, and obtain $s_c = 0.3565(12)$. This value
agrees well with the previous result \cite{farhi12} but has smaller uncertainty.

\begin{figure}
\centerline{\includegraphics[width=8.0cm, clip]{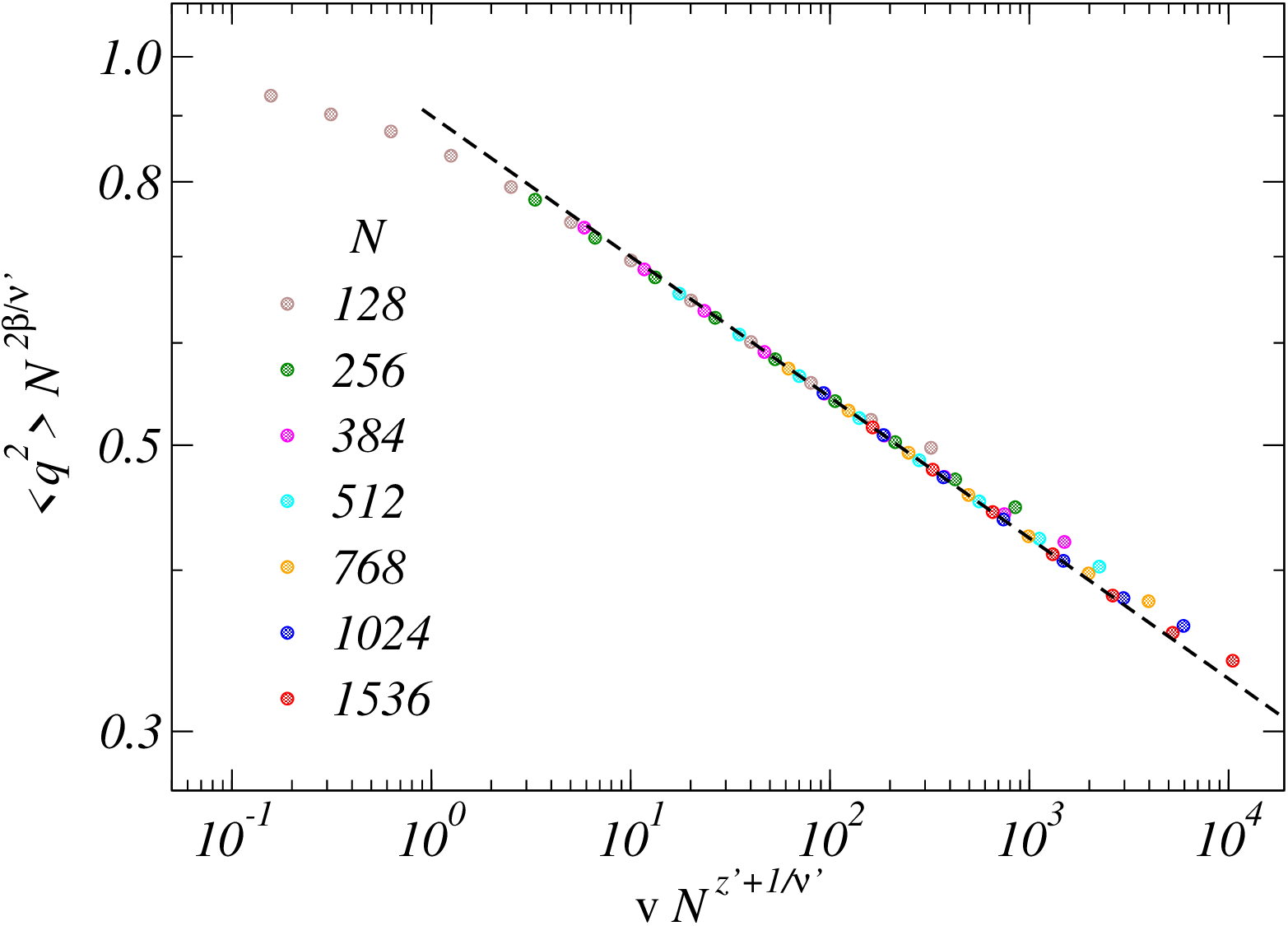}}
\vskip-2mm
\caption{(Color online) Optimized scaling collapse of the order parameter in critical quenches of 3-regular graphs, giving
the exponents listed in the text. The line has slope given in Eq.~(\ref{xdef}) and the points above it (which are excluded in the
fitting procedure) deviate due to high-velocity cross-overs \cite{liu14}.}
\label{fig3}
\vskip-3mm
\end{figure}

Performing additional quenches to the above determined $s_c$, we next extract critical exponents. A scaling graph with data for several system sizes 
is shown in Fig.~\ref{fig3}. Here the exponents are treated as adjustable parameters for obtaining optimal data collapse. After performing an error 
propagation analysis we obtain $\beta/\nu'=0.43 \pm 0.02$ and the KZ exponent $z'+1/\nu'=1.34 \pm 0.11$.

Interestingly, the exponents, in particular the KZ exponent, differ from those obtained using Landau theory \cite{read95} and other methods \cite{miller93} for 
large-$d$ and fully connected ($d=\infty$ \cite{ray89}) Ising models in a transverse field; $\beta=1$, $\nu'=2$ and $z'=1/4$ ($d_u=8$), i.e., $\beta/\nu'= 1/2$ 
and $z'+1/\nu' = 3/4$. In the simplest scenario the 3-regular graphs should have the same exponents. A QMC calculation for the fully-connected
model in Ref.~\onlinecite{alvarez96} was not in complete agreement with the above values. It was argued that $z=4$ ($z'=1/2$ if $d_u=8$), 
$\nu=1/4$ ($\nu'=2$), and $\beta \approx 1$, thus $\beta/\nu' \approx 0.5$ and $z'+1/\nu' = 1$. We have also studied the same fully-connected model, using the 
methods discussed above, and obtained $z'+1/\nu' = 0.83 \pm 0.12$ and $\beta/\nu' = 0.47 \pm 0.03$, in good agreement with the analytical $d=\infty$ values
\cite{read95} (and $s_c$ agrees well with Refs.~\onlinecite{miller93,alvarez96}). A potential source of the disagreement in the case 
of the 3-regular graphs is logarithmic scaling corrections \cite{read95}. However, we do not see any obvious signs of log corrections and the discrepancy in $z'+1/\nu'$ appears 
larger than might be expected from logs alone. Our results therefore suggest other, unknown effects in the $3$-regular graphs.

{\it Implications for quantum computing.}---In the classical 3-regular graphs the KZ exponent is $z'+1/\nu'=1$, while in the quantum system 
$z'+1/\nu' \approx 1.3$. Thus, by Eq.~(\ref{vc}) the adiabatic annealing time grows {\it faster} with $N$ in QA (while in the fully-connected model
it grows slower). Furthermore, since the order parameter scales as $N^{-\beta/\nu'}$, the critical cluster is less dense with QA, i.e., further from 
the solution at $s = 1$ (which applies also to the fully-connected model). Thus, in both these respects QA on 3-regular graphs performs worse than SA in passing 
through the critical boundary into the extended glass phase in the $(h,T)$ plane. While our results do not contain any quantitative information on the process 
continuing from $s_c$ to $s=1$ (where the annealing time may grow exponentially in $N$ \cite{farhi12}), it is discouraging that the important initial stage 
of QA in reaching the glass phase is less efficient than SA. It is known that QA can, in principle, be made more efficient than SA for a given problem by 
changing the quantum term (the driver) \cite{nishimori14,castelnovo05}. However, to make fair comparisons, one should then allow also more complex SA 
evolution, e.g., going beyond just changing $T$.

It would be interesting to study velocity scaling also with the D-Wave device \cite{dwave11,dwave13}, not only with complex frustrated couplings but 
even in simpler cases such as a critical ferromagnet. This would give valuable insights into the annealing process, which in the D-Wave device 
certainly is influenced by temperature effects \cite{dwave11}, in contrast to $T=0$ coherent quantum dynamics studied here.

{\it Acknowledgments.}---We would like to thank Claudio Chamon, David Huse, Subir Sachdev, and Peter Young for stimulating discussions. 
This work was supported by the NSF under grant No.~PHY-1211284. AWS also acknowledges support from the Simons Foundation.

\null

\end{document}